\documentclass[prd,aps,showpacs,amsmath,amssymb,twocolumn]{revtex4}
\usepackage{graphicx}% Include figure files
\usepackage{dcolumn}% Align table columns on decimal point
\usepackage{bm}% bold math

\begin{document}

%%%%
%    Greek Letters
%

\let\a=\alpha      \let\b=\beta       \let\c=\chi        \let\d=\delta
\let\e=\varepsilon \let\f=\varphi     \let\g=\gamma      \let\h=\eta
\let\k=\kappa      \let\l=\lambda     \let\m=\mu
\let\o=\omega      \let\r=\varrho     \let\s=\sigma
\let\t=\tau        \let\th=\vartheta  \let\y=\upsilon    \let\x=\xi
\let\z=\zeta       \let\io=\iota      \let\vp=\varpi     \let\ro=\rho
\let\ph=\phi       \let\ep=\epsilon   \let\te=\theta
\let\n=\nu
\let\D=\Delta   \let\F=\Phi    \let\G=\Gamma  \let\L=\Lambda
\let\O=\Omega   \let\P=\Pi     \let\Ps=\Psi   \let\Si=\Sigma
\let\Th=\Theta  \let\X=\Xi     \let\Y=\Upsilon

%
%%%

%%%
%    Calligraphic letters
%

\def\cA{{\cal A}}                \def\cB{{\cal B}}
\def\cC{{\cal C}}                \def\cD{{\cal D}}
\def\cE{{\cal E}}                \def\cF{{\cal F}}
\def\cG{{\cal G}}                \def\cH{{\cal H}}
\def\cI{{\cal I}}                \def\cJ{{\cal J}}
\def\cK{{\cal K}}                \def\cL{{\cal L}}
\def\cM{{\cal M}}                \def\cN{{\cal N}}
\def\cO{{\cal O}}                \def\cP{{\cal P}}
\def\cQ{{\cal Q}}                \def\cR{{\cal R}}
\def\cS{{\cal S}}                \def\cT{{\cal T}}
\def\cU{{\cal U}}                \def\cV{{\cal V}}
\def\cW{{\cal W}}                \def\cX{{\cal X}}
\def\cY{{\cal Y}}                \def\cZ{{\cal Z}}

\def\dbd{{$0\nu 2\beta\,$}}
%
%%%%

\newcommand{\Ns}{N\hspace{-4.7mm}\not\hspace{2.7mm}}
\newcommand{\qs}{q\hspace{-3.7mm}\not\hspace{3.4mm}}
\newcommand{\ps}{p\hspace{-3.3mm}\not\hspace{1.2mm}}
\newcommand{\ks}{k\hspace{-3.3mm}\not\hspace{1.2mm}}
\newcommand{\des}{\partial\hspace{-4.mm}\not\hspace{2.5mm}}
\newcommand{\desco}{D\hspace{-4mm}\not\hspace{2mm}}

%%%%

%\draft command makes pacs numbers print
%\draft

\title{\boldmath Neutrinoless double beta decay and QCD corrections }

\author{Namit Mahajan
}
\email{nmahajan@prl.res.in}
\affiliation{
 Theoretical Physics Division, Physical Research Laboratory, Navrangpura, Ahmedabad
380 009, India
}

%\date{\today}

\begin{abstract}
We consider one loop QCD corrections and renormalization group running of the neutrinoless double beta decay amplitude 
focusing on the short-range part of the amplitude (without the light neutrino exchange) and find that these corrections can be 
sizeable. Depending on the operator under consideration, 
there can be moderate to large cancellations or significant enhancements. We discuss several specific examples in this context.
Such large corrections will lead to significant shifts in the half-life estimates which currently are known to be plagued with the
uncertainties due to nuclear physics inputs to the physical matrix elements. 
\end{abstract}

\maketitle
%\narrowtext

%\section{Introduction}

It is now experimentally well established that neutrinos have mass and they mix with each other (see \cite{Tortola:2012te} for the best 
fit values of the parameters). 
Being electrically neutral allows the possibility of them to be Majorana particles \cite{Majorana:1937vz}. The observation of neutrinoless 
double beta (\dbd) decay, $(A,Z)\rightarrow (A,Z+2) + 2e^-$, will establish the Majorana nature and lepton number violation beyond any doubt
\cite{Furry:1939qr}.
Therefore, the search for neutrinoless double beta decay continues to be an important area. 
Theoretically as well, \dbd decay is heralded as a useful probe of physics beyond the standard model (SM).
 \dbd can potentially discriminate between the two hierarchies of the neutrino masses, and this, in turn can be used 
 to rule out specific models of neutrino mass generation. In the context of models which involve TeV scale particles, 
 like low scale seesaw models or low energy supersymmetric models including models with R-parity violation, \dbd 
 imposes stringent constraints on the model parameters. The same set of diagrams, with appropriate changes in the momentum flow,
 can lead to interesting signatures at LHC. Constraints from \dbd thus can prove rather useful for phenomenological studies 
 (see e.g. \cite{Keung:1983uu} for an incomplete list discussing various aspects).

 The \dbd decay amplitude can be split into the so called long-range and short-range parts 
 (for a review of theoretical and experimental issues and the sources of uncertainties and errors, 
 see \cite{Rodejohann:2011mu} and references therein ). 
 Here, the long-range refers to the fact that there is an intermediate light neutrino involved. 
 This should be contrasted with the short-range part of the amplitude in which the intermediate particles are all much 
 much heavier that the relevant scale of the process $\sim {\mathcal{O}}$(GeV). In such a case, the heavier degrees 
 of freedom can be systematically integrated out leaving behind a series of operators built out of low energy fields 
 weighted by coefficients, called Wilson coefficients (denoted by $C_i$ below), which are functions of the parameters of 
 the large mass degrees of freedom that have been integrated out (see e.g \cite{Georgi:1994qn}). This provides a very convenient framework to evaluate the 
 decay amplitude in terms of short distance coefficients which encode all the information about the high energy physics one 
 may be trying to probe via a low energy process. This also neatly separates the particle physics input from the nuclear
 physics part which enters via the nuclear matrix elements (NMEs) of the quark level operators sandwiched between the nucleon states. 
 In what follows, the discussion will be centered around the short range part though we believe that many of the arguments and 
 results may also apply to the long range part. More care may be needed in the latter case though.

Given a specific model it is straightforward to write down the amplitude for the quark level 
\dbd process and compute the short distance coefficient. The complete amplitude then involves NMEs. 
At present, the biggest source of uncertainty stems from the NMEs, and theoretical predictions show a marked sensitivity on the
NMEs used (see \cite{Simkovic:2007vu} for some of the recent NME calculations and predictions for \dbd rates).
On the experimental side, 
studies have been carried out on several nuclei. 
Only one of the experiments, the Heidelberg-Moskow collaboration (HM collab.) \cite{KlapdorKleingrothaus:2006ff} has claimed observation of \dbd signal in $^{76}{\mathrm Ge}$. 
The half-life at $68\%$ confidence level is: $T^{0\nu}_{1/2}(^{76}{\mathrm Ge}) = 2.23^{+0.44}_{-0.31}\times 10^{25}\,
{\mathrm yr}$. A combination of the Kamland-Zen \cite{Gando:2012zm} and EXO-200 \cite{Auger:2012ar} results, both using $^{136}{\mathrm Xe}$, 
yields a lower limit on the half-life $T^{0\nu}_{1/2}(^{136}{\mathrm Xe}) > 3.4 \times 10^{25}\, {\mathrm yr}$ 
which is at variance with the HM claim. Very recently GERDA experiment reported the lower limit on the half-life based 
on the first phase of the experiment \cite{Agostini:2013mzu}: $T^{0\nu}_{1/2}(^{76}{\mathrm Ge}) > 2.1 \times 10^{25}\, {\mathrm yr}$. A combination
of all the previous limits results in a lower limit $T^{0\nu}_{1/2}(^{76}{\mathrm Ge}) > 3.0 \times 10^{25}\, {\mathrm yr}$
at $90\%$ confidence level. The new GERDA result (and the combination) is (are) again at odds with the positive claim of HM collab.
The GERDA results have been challenged \cite{Klapdor-Kleingrothaus:2013cja} on account of low statistics and poorer resolution. 
Very clearly, there is some tension among the
experimental results and higher statistics in future will shed more light.
To reduce the dependence (or sensitivity) on NMEs, predictions for \dbd for various nuclei can be compared. Further, it is necessary
to establish if the long-range contribution, coming from the light neutrino exchange, can saturate the experimental limits (or 
positive claims). This is investigated in \cite{Dev:2013vxa}, and the conclusion drawn is that the light neutrino exchange falls short of
saturating the current limits. Also, for some choices of NMEs, the $^{76}{\mathrm Ge}$ positive result can be consistent with 
$^{136}{\mathrm Xe}$ limits when considered individually but not when combined.

In view of the immense importance of \dbd, both experimentally and theoretically, it is important to ensure 
that theoretical calculations are very precise. In the present article, we consider dominant one loop QCD 
corrections and renormalization group effects to the \dbd amplitude. To the best of our knowledge, this has not 
been studied before and as we show below, QCD corrections can have a significant impact on the \dbd rate, thereby 
impacting the constraints on the model parameters.

We begin by recapitulating the essential steps in arriving at the final amplitude for \dbd. 
Using the Feynman rules for a given model, all possible terms can be easily written. 
Since the momentum flowing through any of the internal lines is far smaller than the masses of the 
respective particles and can be neglected, this leads to the low energy amplitude at the quark level. Parts of the amplitude may require Fierz rearrangement (for example in supersymmetric theories) to express it in colour singlet form which can then be sandwiched between the nucleon states after taking the non-relativistic limit. This last step results in NMEs. 
We shall not be concerned with the issue of uncertainties creeping in due to NME calculations here.
We shall, rather, choose to work with a particular set of NMEs and focus on the impact of perturbative 
QCD corrections.  As an example, consider a heavy right handed neutrino and SM gauge group. The resulting amplitude is of the form
\begin{eqnarray}
{\mathcal{A}} &\sim& \frac{1}{M_W^4M_N}\bar{u}\gamma_{\mu}(1-\gamma_5)d\,\bar{e}\gamma^{\mu}\gamma^{\nu}(1+\gamma_5)e^c\,
\bar{u}\gamma_{\nu}(1-\gamma_5)d \nonumber \\
&=&  \underbrace{\frac{1}{M_W^4M_N}}_{G}\underbrace{\bar{u}\gamma_{\mu}(1-\gamma_5)d\,
\bar{u}\gamma^{\mu}(1-\gamma_5)d}_{{\mathcal{J}}_{q,\mu}{\mathcal{J}}_q^{\mu}}\, \underbrace{\bar{e}(1+\gamma_5)e^c}_{j_l}
\end{eqnarray}
where we used $\gamma_{\mu}\gamma_{\nu} = g_{\mu\nu}-2i\sigma_{\mu\nu}$ and the fact that $\bar{e}\sigma_{\mu\nu}(1+\gamma_5)e^c $ 
vanishes identically. So does $\bar{e}\gamma_{\mu}e^c$. This was noted in \cite{Prezeau:2003xn}.
These, thus, restrict the form of the leptonic current. $G$ denotes the analogue of Fermi constant. 
The exact form of $G$ will be
model dependent. The physical \dbd amplitude is written as
\begin{equation}
{\mathcal{A}}_{0\nu 2\beta} = \langle f\vert i{\mathcal{H}}_{\mathrm eff}\vert i\rangle \sim G\,
\underbrace{\langle f\vert {\mathcal{J}}_{q,\mu}{\mathcal{J}}_q^{\mu}\vert i\rangle}_{\boldmath NME}\,j_l
\end{equation}
This clearly illustrates how the short distance or high energy physics separates from the low energy matrix elements. 
The effective Hamiltonian for a given model is expressed as a sum of operators, $O_i$ weighted by the Wilson coefficents $C_i$: 
%\begin{equation}
${\mathcal{H}}_{\mathrm eff} = G_i C_i O_i$, 
%\end{equation}
where we have allowed for more than one $G$ for more complicated theories. In the above case, there is only one operator 
$O_1 = {\mathcal{J}}_{q,\mu}{\mathcal{J}}_q^{\mu}\,j_l = \bar{u_i}\gamma_{\mu}(1-\gamma_5)d_i\,\bar{u_j}\gamma^{\mu}(1-\gamma_5)d_j\,
\bar{e}(1+\gamma_5)e^c$ ($i,j$ denoting the colour indices) and the corresponding Wilson coefficient $C_1=1$. 
In other models like SUSY with R-parity violation \cite{Mohapatra:1986su} or leptoquarks \cite{Hirsch:1996qy}, Fierz transformations have to be employed to bring the 
operators in the colour matched form. The specific NME that finally enters the \dbd rate depends on the Lorentz and Dirac structure 
of the quark level operator involved.

\begin{figure}[ht!]
\vskip 0.32cm
\hskip 1.35cm
\hbox{\hspace{0.03cm}
\hbox{\includegraphics[scale=0.85]{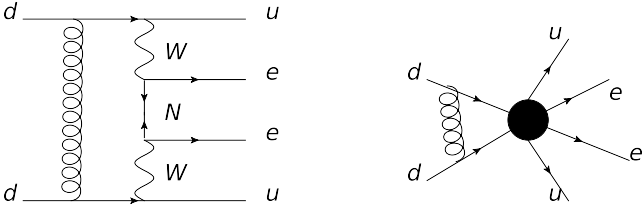}}
}
\caption{Representative Feynman diagrams (drawn using the package JaxoDraw \cite{Binosi:2003yf}) showing one loop QCD corrections.
Left: Full theory and Right: Effective theory
 }
 \label{fig1}
\end{figure}

This is not the entire story. From the effective field theory point of view, the integrating out of the heavier 
degrees of freedom happens at the respective thresholds and then the obtained effective Lagrangian/Hamiltonian has to be properly 
evolved down to the relevant physical scale of the problem ($\sim {\mathcal{O}}$(GeV) in the present case). 
This is similar to what happens in non-leptonic meson decays (see for example \cite{Buchalla:1995vs}). 
For simplicity, we assume that the heavy 
particles are all around the electroweak (EW) scale and in obtaining the numerical values, we shall put $M_W$ as the scale for all.
This facilitates one step integrating out of all the heavy degrees of freedom. Therefore, the above statement about $C_1$ being
unity should now be written as $C_1(M_W) = 1$. Next consider one loop QCD corrections. The full amplitude is evaluated with one 
gluon exchange ($O(\alpha_s) $) and matched with the  amplitude at the same order in $\alpha_s$ in the effective theory. Fig.\ref{fig1}
shows representative diagrams in the full and effective theory. 
This has two effects: (i) $C_1$ gets corrected and reads $C_1(M_W) = 1 + \frac{\alpha_s}{4\pi}{\mathcal{N}} \ln\left(\frac{M_W^2}{\mu_W^2}\right)$, 
where $\mu_W$ is the renormalization scale and ${\mathcal{N}}$ is a calculable quantity. This coefficient is then evolved down to the 
${\mathcal{O}}$(GeV) using the renormalization group (RG) equations; (ii) QCD corrections induce the colour mismatched 
operator $O_2 = \bar{u_i}\gamma_{\mu}(1-\gamma_5)d_j\,\bar{u_j}\gamma^{\mu}(1-\gamma_5)d_i\,\bar{e}(1+\gamma_5)e^c$ with 
coefficient $C_2 = \frac{\alpha_s}{4\pi}{\mathcal{N}}' \ln\left(\frac{M_W^2}{\mu_W^2}\right)$. When evaluating the quark level matrix
element in the effective theory, both the operators contribute and in fact lead to mixing. This approach is a consistent one and also reduces the scale dependence of
the physical matrix elements. Without following the above steps, the short distance coefficient would have been evaluated
at the high scale while the physical matrix elements at a low scale, leading to large scale dependence, which is not a
physical effect but rather an artifact of the calculation.

Armed with this machinery, we now consider specific examples to bring out the impact of QCD corrections. 
As mentioned above, to simplify the discussion, we assume all the heavy particles beyond the SM to be around the EW scale. 
The technical details and explicit expressions for some of the models leading to neutrinoless double beta decay and 
related phenomenology will be presented elsewhere. Here we provide approximate numerical values of the Wilson coefficients 
of the operators considered. For the time being, we neglect the mixing of operators under renormalization. This can have a large
impact on some of the coefficients but their inclusion is beyond the scope of the present work.

First we consider left-right symmetric model and focus our attention on operators generated due to $W_L$ and $W_R$ exchange:
\begin{eqnarray}
O^{LL}_1 &=& \bar{u_i}\gamma_{\mu}(1-\gamma_5)d_i\,\bar{u_j}\gamma^{\mu}(1-\gamma_5)d_j\,\bar{e}(1+\gamma_5)e^c \nonumber \\
O^{LL}_2 &=& \bar{u_i}\gamma_{\mu}(1-\gamma_5)d_j\,\bar{u_j}\gamma^{\mu}(1-\gamma_5)d_i\,\bar{e}(1+\gamma_5)e^c \nonumber \\
O^{RR}_1 &=& \bar{u_i}\gamma_{\mu}(1+\gamma_5)d_i\,\bar{u_j}\gamma^{\mu}(1+\gamma_5)d_j\,\bar{e}(1+\gamma_5)e^c \nonumber \\
O^{RR}_2 &=& \bar{u_i}\gamma_{\mu}(1+\gamma_5)d_j\,\bar{u_j}\gamma^{\mu}(1+\gamma_5)d_i\,\bar{e}(1+\gamma_5)e^c \nonumber \\
O^{LR}_1 &=& \bar{u_i}\gamma_{\mu}(1-\gamma_5)d_i\,\bar{u_j}\gamma^{\mu}(1+\gamma_5)d_j\,\bar{e}(1+\gamma_5)e^c \nonumber \\
O^{LR}_2 &=& \bar{u_i}\gamma_{\mu}(1-\gamma_5)d_j\,\bar{u_j}\gamma^{\mu}(1+\gamma_5)d_i\,\bar{e}(1+\gamma_5)e^c
\end{eqnarray}
Following the general steps outlined above, the Wilson coefficents can be evaluated at the high scale and run down to
$\mu \sim {\mathcal{O}}$(GeV) (see also \cite{Cho:1993zb}). Their approximate values read:
\begin{eqnarray}
C^{LL,RR}_1 \sim 1.3 &,& C^{LL,RR}_2 \sim -0.6 \nonumber\\
C^{LR,RL}_1 \sim 1.1 &,& C^{LR,RL}_2 \sim 0.7
\end{eqnarray}
To evaluate the physical matrix elements, the colour mismatched operators $O^{AB}_2$ have to be Fierz transformed. 
Under Fierz rearrangement, $(V-A)\otimes (V-A)$ and $(V+A)\otimes (V+A)$ retain their form while
$(V-A)\otimes (V+A) \rightarrow -2 (S-P)\otimes (S+P)$. With this rearrangement, $LL,\,RR$ operators effectively 
yield $C^{LL,RR}_1+C^{LL,RR}_2$ as the effective couplings with the same NMEs involved, implying substantial cancellation 
(by about a factor of two). The $LR$ operator Fierz transformed brings in a different combination of NMEs. 
Explicitly, following for example the last reference in  \cite{Rodejohann:2011mu}, 
we have the following (not showing the lepton current explicitly):
\begin{equation}
\langle {\mathcal{J}}^{(V\pm A)}{\mathcal{J}}_{(V\pm A)}\rangle \propto \frac{m_A}{m_Pm_e} 
({\mathcal{M}}_{GT,N}\,  \mp \alpha^{SR}_3 {\mathcal{M}}_{F,N})
\end{equation}
where $\vert{\mathcal{M}}_{GT,N}\vert \sim (2-4)\vert{\mathcal{M}}_{F,N}\vert$
for all the nuclei considered, and $\alpha^{SR}_3 \sim 0.63$. Thus, to a good accuracy the above matrix element is 
essentially governed by ${\mathcal{M}}_{GT,N}$. In the above equation, the relative negative sign between the two terms
on the right hand side corresponds to $(V+A)\otimes (V+A)$ and $(V-A)\otimes (V-A)$ structures on the left hand side, 
while for the $(V+A)\otimes (V-A)$ structure, the relative sign is positive.

On the other hand, 
\begin{equation}
\langle {\mathcal{J}}^{(S\pm P)}{\mathcal{J}}_{(S\pm P)}\rangle \propto - \alpha^{SR}_1 {\mathcal{M}}_{F,N}
\end{equation}
with $\alpha^{SR}_1 \sim 0.145\frac{m_A}{m_Pm_e}$.
Clearly, the Fierz transformed operator in this case turns out to be subdominant. This simple exercise illustrates 
the large impact and importance of QCD corrections in the context of \dbd. As obtained above, QCD corrections can lead to 
substantial shift in the \dbd rate for specific models, thereby changing the limits on the model parameters significantly.

As our next example, consider theories where the interactions are $S\pm P$ form, like SUSY with R-parity violation or
leptoquarks etc. In such cases, the operators have the structure:
\begin{eqnarray}
O^{SP\pm\pm}_1 &=& \bar{u_i}(1\pm\gamma_5)d_i\, \bar{u_j}(1\pm\gamma_5)d_j\, \bar{e}(1+\gamma_5)e^c \nonumber \\
O^{SP\pm\pm}_2 &=& \bar{u_i}(1\pm\gamma_5)d_j\, \bar{u_j}(1\pm\gamma_5)d_i\, \bar{e}(1+\gamma_5)e^c \nonumber \\
O^{SP+-}_1 &=& \bar{u_i}(1+\gamma_5)d_i\, \bar{u_j}(1-\gamma_5)d_j\, \bar{e}(1+\gamma_5)e^c \nonumber \\
O^{SP+-}_2 &=& \bar{u_i}(1+\gamma_5)d_j\, \bar{u_j}(1-\gamma_5)d_i\, \bar{e}(1+\gamma_5)e^c 
\end{eqnarray}
The Wilson coefficients of the colour mismatched operators are about 0.1-0.5 times those of the colour allowed operators in magnitude.
This could be argued from the $1/N_c(\,\sim 0.3\,\,\textrm{for} N_c=3)$ counting rules for the colour mismatched structures, up to factors of order unity. 
Following the same chain of arguments, the colour mismatched operators need to be Fierz transformed before computing the physical
matrix elements. Under Fierz transformations we have: $(S+P)\otimes (S-P) \rightarrow \frac{1}{2}(V+A)\otimes (V-A)$ implying that
the colour mismatched operator, after Fierz transformation, may provide the dominant contribution (see Eq.(5) and Eq.(6)). 
Consequently the amplitudes, and therefore the limits on parameters may change by a factor of five or so. That the colour mismatched
operator can provide a large contribution is again something we are familiar with from $K\to\pi\pi$ decays where the QCD (and
electroweak) penguin operator after Fierz transformation gives the dominant contribution, though QCD and electroweak penguin contributions
tend to cancel each other in this case.

The most interesting and the largest effect in the examples considered above comes about when considering the $O^{SP++,--}$ operators. 
$(S\pm P)\otimes (S\pm P) \rightarrow \frac{1}{4}[2(S\pm P)\otimes (S\pm P) - (S\pm P)\sigma_{\mu\nu}\otimes\sigma^{\mu\nu}]$ 
under Fierz rearrangement. The tensor-pseudotensor structure yields the following NME:
\begin{equation}
\langle {\mathcal{J}}^{\mu\nu}{\mathcal{J}}_{\mu\nu}\rangle \propto -\alpha^{SR}_2 {\mathcal{M}}_{GT,N}
\end{equation}
with $\alpha^{SR}_2 \sim 9.6\frac{m_A}{m_Pm_e}$ which is about $200$ times larger than 
$\langle {\mathcal{J}}^{(S\pm P)}{\mathcal{J}}_{(S\pm P)}\rangle$. Conservatively taking the corresponding 
Wilson coefficient to be $0.1$ of the colour allowed operator, the relative contributions are:
\begin{equation}
\vert\frac{O^{SP++}_2}{O^{SP++}_1}\vert \geq 10
\end{equation}
The above discussion makes it very clear that the QCD corrections to \dbd are rather important and should be included 
systematically. 
These corrections can be as large as or in fact larger than in most cases than the uncertainty due to NMEs 
and are independent of the particular set of NMEs considered.
As eluded to above, we have considered only pairs of operators $O^{AB}_1,\, O^{AB}_2$ while obtaining the approximate values 
of $C's$ at the low scale. The effect of mixing with other operators has been ignored at this stage. This could further lead
to significant corrections for some of the operators. We plan to systematically investigate these issues elsewhere.
This (and the shift above) is rather large and can completely change the phenomenological constraints. 
In theories with many contributions to \dbd, it is essential to understand the interplay between different competing
amplitudes to set limits on the couplings and masses of the particles. In such cases, the discussion above becomes even more
important.
Low (TeV) scale models appear to be attractive due to plausible signatures at LHC, where QCD corrections will be
inevitable. It is therefore important to include the dominant QCD corrections at the very least in order to set meaningful
limits on model parameters. \\

In this article we have investigated the impact of one loop QCD corrections to the \dbd amplitude. 
This, to the best of our knowledge, is the first time this issue has been discussed. We found that QCD corrections can
have a large impact, ranging from near cancellation to huge enhancement of the \dbd rate. Since \dbd is an important 
process to search experimentally and has the potential to link seemingly unrelated processes, particularly in the context
of TeV scale models, it is rather important to ensure that theoretical predictions are precise enough to be compared to
the experimental results. As such, the calculations suffer from large uncertainties due to the choice of NMEs, 
which are non-perturbative in nature. What we have found is that even perturbative corrections have the potential to 
shift the predictions by a large amount. This by itself is a rather important aspect and such corrections need to 
be systematically computed for various models of interest. The shift in the limits on model parameters also implies 
that the related phenomenology at say the LHC (in specific models) will also get modified. There are other issues related 
to operator mixing which have not been incorporated here. These may also become important in the context of specific theories
and should be consistently included. Furthermore, QCD corrections need to be evaluated for the light neutrino exchange contribution
as well.
As mentioned in the beginning, the light neutrino contribution is unable to saturate the present experimental limits. It remains to be seen if 
including the radiative corrections eases out this tension, and to what extent \cite{nm}.

%expt prospect
%\vskip 3cm

%MFV

%

\end{document}